\documentclass[10pt]{article}

\usepackage{amsmath}
\usepackage{amssymb}
\usepackage{graphics}
\usepackage{rotating}
\usepackage{cite}
\usepackage{color}
\usepackage{fancybox}
\usepackage{pstricks}
\usepackage{xcolor}
\usepackage{shadowtext}
\usepackage{multicol}

\def\0{\mbox{\tiny $0$}}
\def\1{\mbox{\tiny $1$}}
\def\2{\mbox{\tiny $2$}}
\def\3{\mbox{\tiny $3$}}
\def\4{\mbox{\tiny $4$}}
\def\5{\mbox{\tiny $5$}}
\def\6{\mbox{\tiny $6$}}
\def\7{\mbox{\tiny $7$}}
\def\8{\mbox{\tiny $8$}}
\def\9{\mbox{\tiny $9$}}
\definecolor{navy}{rgb}{0,0,.6}
\definecolor{jour}{rgb}{0,0.6,.4}
\definecolor{jbul}{rgb}{0.7,0.,.4}
\begin{document}
%
\thispagestyle{empty}
\setcounter{page}{0}

\begin{center}
\shadowrgb{0.8,0.8,1}
\shadowoffset{4pt}
\shadowtext{
\color{navy}
\fontsize{18}{18}\selectfont
\bf THE ASYMMETRIC GOOS-H\"ANCHEN EFFECT}
\end{center}

\vspace*{1cm}

\begin{center}
\shadowrgb{0.8, .8, 1}
\shadowoffset{2.5pt}
\shadowtext{\color{jbul}
\fontsize{13}{13}\selectfont
$\boldsymbol{\bullet}$}
\shadowtext{\color{jour}
\fontsize{15.5}{15.5}\selectfont
\bf Journal of Optics 16, 015702-7 (2014)}
\shadowtext{\color{jbul}
\fontsize{13}{13}\selectfont
$\boldsymbol{\bullet}$}
\end{center}

\vspace*{1cm}

\begin{center}
\begin{tabular}{cc}
\begin{minipage}[t]{0,5\textwidth}
{\bf Abstract}.
We show in which conditions optical gaussian beams, propagating  throughout  an homogeneous dielectric right angle prism, present an {\em asymmetric} Goos-H\"anchen (GH) effect. This asymmetric behavior is seen for incidence at critical angles and happens in the propagation direction of the outgoing beam.   The asymmetric GH effect can be also seen as an {\em amplification} of the standard  GH shift.  Due to the fact that it only depends on the ratio between the wavelength and the minimal waist size of the incoming gaussian beam, it  can be also used to determine one of these parameters. Multiple peaks interference is an additional phenomenon seen  in the presence of such asymmetric effects.
\end{minipage}
& \begin{minipage}[t]{0,5\textwidth}
{\bf Manoel P. Araujo}\\
Gleb Wataghin Physics Institute\\
State University of Campinas (Brazil)\\
{\color{navy}{{\bf mparaujo@ifi.unicamp.br}}}
\hrule
\vspace*{0.2cm}
{\bf Silv\^ania A. Carvalho}\\
Department of Applied Mathematics\\
State University of Campinas (Brazil)\\
{\color{navy}{{\bf  silalves@ime.unicamp.br}}}
\hrule
\vspace*{0.2cm}
{\bf Stefano De Leo}\\
Department of Applied Mathematics\\
State University of Campinas (Brazil)\\
{\color{navy}{{\bf  deleo@ime.unicamp.br}}}
\end{minipage}
\end{tabular}
\end{center}

\vspace*{2cm}

\begin{center}
\shadowrgb{0.8, 0.8, 1}
\shadowoffset{2pt}
\shadowtext{\color{navy}
{\bf
\begin{tabular}{ll}
I. &INTRODUCTION \\
II. &DIELECTRIC SYSTEM GEOMETRY AND OUTGOING BEAM \\
III. & ASYMMETRIC GH EFFECT AND MULTIPLE PEAKS INTERFERENCE \\
IV.  & CONCLUSIONS \\
& \\
& \,[\,13 pages, 6 figures\,]
\end{tabular}
}}
\end{center}

\vspace*{7cm}

\begin{flushright}
\shadowrgb{.8, .8, 1}
\shadowoffset{2pt}
\shadowtext{\color{jbul}
\fontsize{11}{11}\selectfont
$\boldsymbol{\bullet}$}
\hspace*{-.2cm}
\shadowtext{
\color{jour}
\fontsize{15.5}{15.5}\selectfont
$\boldsymbol{\Sigma\hspace*{0.06cm}\delta\hspace*{0.035cm}\Lambda}$}
\hspace*{-.35cm}
\shadowtext{\color{jbul}
\fontsize{11}{11}\selectfont
$\boldsymbol{\bullet}$}
\end{flushright}




\newpage


\section*{\normalsize I. INTRODUCTION}

The behavior of laser gaussian beams in the presence of {\em symmetric} and {\em asymmetric} wave number distributions is the subject matter of this paper. The importance of the difference among
such kinds of distributions can be understood if one takes into account the analogy between optics\cite{born,saleh} and non relativistic quantum mechanics\cite{cohen,grif} and discuss the behavior between {\em stationary} and {\em dynamical} maxima.

A non-relativistic free (gaussian) particle in its rest frame is described by the following wave packet
\begin{eqnarray}
\Psi(x,y,t) & = &\Psi_{\0}\,
\frac{d}{4\, \pi}\,\,
\int^{^{+\infty}}_{_{-\infty}}\hspace*{-.5cm}
\mbox{d}k_x\,\int^{^{+\infty}}_{_{-\infty}}\hspace*{-.5cm}
\mbox{d}k_y
\,\, \exp \left[-\frac{
(k_{x}^{^2}+k_{y}^{^2})\,d^{^{2}}}{4}\,+\,i\,\left(\,
k_x\,x +
k_y\,y - \,\hbar\, \frac{k_{x}^{^2}+k_{y}^{^2} }{2\,m}\,\,t\,\right)\right]\nonumber \\
  & = &  \Psi_{\0}\,\,  \mathcal{G}\left(\frac{x}{d}\,,\,  \frac{\hbar\,t}{m\,d^{^{\,2}}}  \right)\, \mathcal{G}\left(\frac{y}{d}\,,\,  \frac{\hbar\,t}{m\,d^{^{\,2}}}  \right)\,\,,
\end{eqnarray}
where
\[\mathcal{G}\left(\alpha\,,\,\beta\right) =  \exp\left[- \,\frac{\alpha^{\2}}{1 + 2\,i\,\beta} \right] \mbox{\huge $/$}  \sqrt{1 + 2\,i\,\beta}\,\,.
\]
This wave convolution is solution of the two-dimensional Schr\"odinger equation\cite{cohen},
\begin{equation}
\label{scr}
\left[\,\partial_{xx}  +\partial_{yy} + 2\, i\,\mbox{$\frac{m}{\hbar}$}\, \partial_t\,\right]\,\Psi(x,y,t)
=0\,\,.
\end{equation}
The gaussian probability density, $\left| \Psi(x,y,t)\right|^{^{2}}$, grows with the beam diameter as  a function of time. Its  maximum, which decreases for increasing values of time, is {\em always}  located at $x=y=0$. It represents a {\em stationary} maximum. It is obvious that the previous analysis is a consequence of the choice of a {\em symmetric} momentum distribution centered in $k_x=k_y=0$.

To illustrate the idea behind our study, let us consider a gaussian momentum distribution with only positive momentum values for $k_y$, i.e.
\begin{eqnarray}
\Phi(x,y,t) & = & \Phi_{\0}\,\,\mathcal{G}\left(\frac{x}{d}\,,\,  \frac{\hbar\,t}{m\,d^{^{\,2}}}  \right)\,
\frac{d}{\sqrt{\pi}}\,\,
\int^{^{+\infty}}_{_{0}}\hspace*{-.5cm}
\mbox{d}k_y
\,\, \exp \left[-\frac{
k_{y}^{^2}\,d^{^{2}}}{4}\,+\,i\,
k_y\,y - \,i\,\hbar \frac{k_{y}^{^2}}{2\,m}\,t\,\right]\nonumber \\
  & = &  \Phi_{\0}\, \, \mathcal{G}\left(\frac{x}{d}\,,\,  \frac{\hbar\,t}{m\,d^{^{\,2}}}\right)\,
   \mathcal{G}\left(\frac{y}{d}\,,\,  \frac{\hbar\,t}{m\,d^{^{\,2}}}\right)
   \,\left[\,1 +\, \mbox{erf}\left(i\,\frac{y}{d}\,\,\mbox{\huge $/$}\,\sqrt{1+\,i\,
 \frac{2\,\hbar\,t}{m\,d^{^{\,2}}}}\, \right) \,\right] \,\,.
\end{eqnarray}
The maximum of this distribution can be estimated  by using a basic principle of asymptotic analysis\cite{dingle}. For oscillatory integrals, rapid oscillations over the range of integration means that the integrand averages to zero. To avoid this cancelation rule, the phase has to be calculated   when  it is stationary, i.e.
\[
0=\left\{\frac{\partial}{\partial k_y} \left[  k_y\,y\,-\,\hbar\,\frac{k_{y}^{^2}}{2\,m}\,t\,\right]\right\}_{_{k_y=\langle k_y\rangle}}\,\,\,\,\,\,\,\Rightarrow\,\,\,\,\,\,\,\,\,\,\,\,\,\,y=\frac{\hbar\,\langle k_{y}\rangle }{m}\,\,t\,\,.
\]
The {\em breaking} of symmetry in the gaussian momentum distribution implies now an expected  value of $k_y$ different from zero,
\begin{equation}
\langle k_y \rangle = \int^{^{+\infty}}_{_{0}}\hspace*{-.5cm}
\mbox{d}k_y\,\,k_y\, \exp \left[-\,\frac{
k_{y}^{^2}\,d^{^{\,2}}}{4}\right]\,\,\mbox{\huge $/$}\, \int^{^{+\infty}}_{_{0}}\hspace*{-.5cm}
\mbox{d}k_y\,\, \exp \left[-\,\frac{
k_{y}^{^2}\,d^{^{\,2}}}{4} \right] = \frac{2}{d\,\sqrt{\pi}}\,\,,
\end{equation}
and, consequently, a {\em dynamical} maximum at
\begin{equation}
y_{max}= \frac{2\,\hbar}{m\,d\,\sqrt{\pi}}\,\,t\,\,.
\end{equation}
For non-relativistic quantum particles, the difference between stationary and dynamical maxima can be roughly represented as the difference between symmetric and asymmetric wave number distributions. The aim of this paper is to investigate in which conditions we can reproduce dynamical maxima for laser gaussian beams propagating throughout an homogeneous dielectric right angle prism.

\noindent
The Maxwell equations
\begin{equation}
\left[ \nabla^{^{2}}-\frac{\partial_{tt}}{c^{^{2}}}\right] E(\boldsymbol{r},t)=0\,\,,
\end{equation}
 for time harmonic electric fields ($\exp[\,-\,i\,\omega \,t\,]$)  and for plane waves, modulated by a complex amplitude  $A(\boldsymbol{r})$, which  travel along the $z$-direction ($\exp[\,i\,k\,z\,]$ with $k=\omega/c=2\,\pi/\lambda$),
\[ E(\boldsymbol{r},t)  =    E_{\0}\,e^{\,i\,(k\,z - \omega\,t)}\,A(\boldsymbol{r})\,\,,\]
 reduce to\cite{arf}
 \begin{eqnarray}
 \label{para0}
0 & =& E_{\0}\,e^{\,i\,(k\,z - \omega\,t)} \left[ \partial_{xx} + \partial_{yy} \partial_{zz} + 2\,i\,k\,\,\partial_z - k^{^{2}}  + \frac{\omega^{\2}}{c^{^{2}}} \right] A(\boldsymbol{r}) \nonumber \\
 & =  & E_{\0}\,e^{\,i\,(k\,z - \omega\,t)} \left[\, \partial_{xx} + \partial_{yy} + \partial_{zz} + 2\,i\,k\,\,\partial_z  \right] \,A(\boldsymbol{r})\,\,.
 \end{eqnarray}
 In the paraxial approximation\cite{born}, $A(\boldsymbol{r})$ is a slowly varying function of $z$ and the previous equation becomes
 \begin{equation}
 \label{para}
\left[\, \partial_{xx} + \partial_{yy} + 2\,i\,k\,\,\partial_z \right] \,A(x,y,z) =0\,\,.
 \end{equation}
 The analogy between the paraxial approximation of the Maxwell equations, Eq.(\ref{para}), and the non-relativistic Sch\"odinger equation,  Eq.(\ref{scr}),   is then clear if we consider the following correspondence rules\cite{longhi,del1}
 \[z\longleftrightarrow t\,\,\,\,\,\,\,\mbox{and}\,\,\,\,\,\,\,k\longleftrightarrow m/\hbar\,\,. \]
 It is interesting  to ask in which circumstances, by using optical paraxial beams, it is possible to have  an asymmetrical wave number distribution and, consequently, produce a dynamical shift. The study presented in this paper aims to give a satisfactory answer to this intriguing question.

 Gaussian beams are the simplest type of paraxial beams provided by a laser source. The electric field amplitude of the  incident  paraxial gaussian beam is given by\cite{del2,del3}
\begin{eqnarray}
\label{incgb}
 E(\boldsymbol{r},t)  &= &   E_{\0}\,e^{\,i\,(k\,z-\omega\,t)}\,\frac{\mbox{w}_{\0}^{\2}}{4 \pi}\,\,   \int^{^{+\infty}}_{_{-\infty}}\hspace*{-.5cm}
\mbox{d}k_x\,\int^{^{+\infty}}_{_{-\infty}}\hspace*{-.5cm}
\mbox{d}k_y\,\, \exp \left[-\frac{(k_{x}^{^2} +
k_{y}^{^2})\,\mbox{w}_{\0}^{\2}}{4} \,+\,i\,\left(\,
 k_x\,x+k_y\,y - \frac{k_{x}^{^2} +
k_{y}^{^2}}{2\,k}\,z
\,\right)
\right]\nonumber \\
 & = &   E_{\0}\,e^{\,i\,(k\,z-\omega\,t)}\,A(\boldsymbol{r})\,\,,
\end{eqnarray}
where $\mbox{w}_{\0}$ is the minimal waist size of the beam. After performing the $k_x$ and $k_y$ integrations,  we obtain
\begin{equation}
A(\boldsymbol{r}) =
\mathcal{G}\left(\frac{x}{\mbox{w}_{\0}}\,,\,\frac{z}{k\,\mbox{w}^{\2}_{\0}} \right)\, \mathcal{G}\left(\frac{y}{\mbox{w}_{\0}}\,,\,\frac{z}{k\,\mbox{w}^{\2}_{\0}} \right) \,\,.
\end{equation}
The density probability distribution of the gaussian electric field,
\begin{equation}
\label{int}
\left|\,E(\boldsymbol{r},t)\,\right|^{^{2}}= \left|\,E_{\0}\,\right|^{^{2}}\,\left[\,\frac{\mbox{w}_{\0}}{\mbox{w}(z)}\,\right]^{^{2}}\exp \left[\,-\,2\,\,\frac{x^{\2}+y^{\2}}{w^{^{2}}(z)}\,\right]\,\,,
\end{equation}
growths in beam diameter as a function of the $z$-distance from the beam waist $\mbox{w}_{\0}$,
\[ \mbox{w}(z) = \mbox{w}_{\0}\,\sqrt{1+ \left(\,\frac{2\,z}{k\,\mbox{w}_{\0}}\,\right)^{^{2}}}\,\,.
\]
The maximum, which decreases for increasing values of $z$, is always located at $x=y=0$. This maximum plays the role of the {\em stationary} maximum for the quantum non-relativistic particle in its rest frame.

In this paper, we investigate the behavior of optical gaussian beams which propagates through a
right angle prism, see Fig.\,1. For incidence angle $\theta>\theta_c$, the beam is totally reflected at the second interface, its wave number distribution is {\em symmetric} and centered at $k_x=k_y=0$.  Consequently,   the Goos-H\"anchen shift\cite{GH}   is {\em stationary} in the direction of the beam propagation.
The optical phenomenon in which linearly polarized light undergoes a small phase shift, $\delta\approx \lambda$,   when totally internally reflected is widely investigated in litterature\cite{GH0,GH1,GH2,GH3,GH4,GH5,GH6,GH7}. For incidence at and  near critical angles\cite{GHc1,GHc2,GHc3}, we find a frequency crossover in the GH shift which leads to an amplification effect, $\delta_c\approx \sqrt{k\,\mbox{w}_{\0}}\,\,\lambda$ .
 In this paper, we shall present a {\em new} effect
for incidence at critical angles.  Depending on the magnitude  of  $k\mbox{w}_{\0}$, {\em only} the positive values of $k_y$, in the wave number distribution, contribute to reflection and this {\em asymmetry} produces a {\em dynamical } Goos-H\"anchen shift. It is thus  the breaking of the symmetry in the wave number distribution  which opens the door to  a dynamical maximum. A detailed analysis of  this new phenomenon will be discussed in section III.  Before of our numerical study, in section II, we introduce our notation and the geometry of the dielectric system used in this paper. In such a section, we also give, for $s$ and $p$ polarized waves, the reflection and transmission coefficients at each interface. Our final considerations and proposals are drawn in the last section.


\section*{\normalsize II. DIELECTRIC SYSTEM GEOMETRY AND OUTGOING BEAM}
 The incident gaussian beam (\ref{incgb}) propagates along the $z$-axis and forms an angle $\theta$ with  $z_{_{in}}$, normal  to the first air/dielectric interface (see Fig.\,1a),
\begin{equation}
\left( \begin{array}{c}
y_{_{in}} \\
z_{_{in}} \end{array} \right) = \left( \begin{array}{rr}
\cos\theta & \sin\theta \\
-\sin\theta & \cos\theta \end{array} \right)\left( \begin{array}{c}
y \\
z\end{array} \right)=  R\left(\theta\right) \left( \begin{array}{c}
y \\
z\end{array} \right)\,\,.
\end{equation}
Observing that the spatial phase of the incoming beam is
\begin{equation}
\boldsymbol{k_{_{in}}}\cdot \,\,\boldsymbol{r_{_{_{in}}}} = \,\,\,\boldsymbol{k}\,\,\cdot \,\,\boldsymbol{r}\,\,,
\end{equation}
with $k_z = k - (k_x^{^2}+k_y^{^2}\,)/2\,k$, we obtain,  for the beam propagating within the dielectric after the first air/dielectric interface, the following phase
\begin{equation} \boldsymbol{q}_{_{in}}\cdot \,\,\boldsymbol{r_{_{_{in}}}} = \,k_x\, x+k_{y_{_{in}}}\, y_{_{in}} + \sqrt{n^{\2}k^{^{2}} - k_x^{^{2}} - k_{y_{_{in}}}^{^{2}}}\,\,z_{_{in}}
\,\,.\end{equation}
In order to follow the beam motion within the dielectric, we have to introduce two new axes rotations, see Fig.\,1a,
\begin{equation}
 \left( \begin{array}{c}
y_{_{out}} \\
z_{_{out}} \end{array} \right) =  R\left(-\frac{3\,\pi}{4}\right) \left(
\begin{array}{c}
y_{_{*}} \\
z_{_{*}} \end{array} \right) =  R\left(-\frac{\pi}{2}\right) \left(
\begin{array}{c}
y_{_{in}} \\
z_{_{in}} \end{array} \right)\,\,,
\end{equation}
with $z_{_{*}}$ and  $z_{_{out}}$  respectively normal  to the second and third  dielectric/air interface. The spatial phase of the beam moving within the dielectric in the direction of the last  dielectric/air discontinuity can be given in terms of the outgoing axes,
\begin{equation}
\label{e13}
 \boldsymbol{q}_{_{out}}\cdot \,\,\boldsymbol{r_{_{_{out}}}}=\,\,\,k_x\, x+q_{y_{_{out}}}\, y_{_{out}} +q_{z_{_{out}}}\, z_{_{out}} \,\,,
\end{equation}
where
\[ \left( \begin{array}{c}
q_{y_{_{out}}} \\
q_{z_{_{out}}} \end{array} \right) =  R\left(-\,\frac{3\,\pi}{4}\right) \left(
\begin{array}{r}
q_{y_{_{*}}} \\
-\,q_{z_{_{*}}} \end{array} \right)=\left( \begin{array}{r}
-\,k_{y_{_{in}}} \\
q_{z_{_{in}}} \end{array} \right) \,\,.
\]
Observe that the spatial phase of the reflected beam at the second dielectric/air interface is  obtained replacing $q_{z_*}$ by $ -\, q_{z_*}$. Finally,
\begin{eqnarray}
\label{e14}
\boldsymbol{k_{_{out}}}\cdot \,\,\boldsymbol{r_{_{_{out}}}} & = & \,\,k_x\, x+q_{y_{_{out}}}\, y_{_{out}} + \sqrt{k^{^{2}} - k_x^{^{2}} - q_{y_{_{out}}}^{^{2}}}\,\,z_{_{out}}
\nonumber \\
& = &\,\,
k_x\, x +k_{y_{_{in}}}\, z_{_{in}} +  k_{z_{_{in}}}\,\,y_{_{in}} \nonumber \\
 &= & \,\, k_x\,x \,+\,  \left[\,k_z\,\cos
(2\,\theta)-k_y\,\sin (2\,\theta)\,\right]\,y \,+\,\,
\left[\,k_z\,\sin (2\,\theta)+k_y\,\cos
(2\,\theta)\,\right]\,z\,\,,
 \end{eqnarray}
As expected from the Snell law\cite{born,saleh},
\begin{equation}
\left[\,\nabla\,(\,\boldsymbol{k_{_{out}}}\cdot \,\,\boldsymbol{r_{_{_{out}}}}\,)\,\right]_{_{(k_x=0,k_y=0)}} =
   [\,0\,,\,k\,\cos(2\,\theta)\,,\,k\,\sin(2\,\theta)\,]\,\,.
\end{equation}
The amplitude of the outgoing beam is  given by\cite{del2,del3}
\begin{equation}
A_{_{out}}^{^{[s,p]}}(\boldsymbol{r},\theta) =  \frac{\mbox{w}_{\0}^{\2}}{4 \pi}\,\,
\int^{^{+\infty}}_{_{-\infty}}\hspace*{-.5cm}
\mbox{d}k_x\,\int^{^{+\infty}}_{_{-\infty}}\hspace*{-.5cm}
\mbox{d}k_y
\,\,T_{_{\theta}}^{^{[s,p]}}(k_x,k_y)\, \exp \left[-\frac{(k_{x}^{^2} +
k_{y}^{^2})\,\mbox{w}_{\0}^{\2}}{4}\,+\,i\,\varphi_{_{out}}(k_x,k_y;\boldsymbol{r},\theta)\,\right]\,\,,
\end{equation}
where
\[
 \varphi_{_{out}}(k_x,k_y;\boldsymbol{r},\theta)=  k_x\,x+k_y\,[\,\cos
(2\,\theta)\,z - \sin (2\,\theta)\,y\,] - \frac{k_{x}^{^2} +
k_{y}^{^2}}{2\,k}\,[\,\cos
(2\,\theta)\,y + \sin (2\,\theta)\,z\,]\]
and $T_{_{\theta}}^{^{[s,p]}}(k_x,k_y)$ are obtained by calculating the reflection and transmission coefficients at each interface. For $s$-polarized waves, this means
\begin{equation*}
 \frac{2\,k_{z_{_{in}}}}{k_{z_{_{in}}} +\,
q_{z_{_{in}}}}\,\,\times\,\,\frac{q_{z_{_{*}}}-k_{z_{_{*}}} }{q_{z_{_{*}}} +\,
k_{z_{_{*}}}}\,\,\exp[\,2\,i\,q_{z_{_{*}}}\, a_{_{*}}]\,\,\times\,\,\frac{2\,q_{z_{_{out}}}}{q_{z_{_{out}}} +\,
k_{z_{_{out}}}}\,\,\exp[\,i\,(q_{z_{_{out}}} - k_{z_{_{out}}})\, a_{_{out}}]\,\,.
\end{equation*}
By using the geometry of the dielectric system, see Fig\,1a, and  Eqs.(\ref{e13},\ref{e14}), we have
\[
a_{_*}=a/\sqrt{2}\,\,\,,\,\,\,\,\,a_{_{out}}=b-a\,\,\,,\,\,\,\,\,q_{z_{_{out}}}=q_{z_{_{in}}}\,\,\,\,\,
\mbox{and}\,\,\,\,\,k_{z_{_{out}}}=k_{z_{_{in}}}\,\,.\]
Consequently, the transmission coefficient becomes
\begin{equation}
T_{_{\theta}}^{^{[s]}}(k_x,k_y)= \frac{4\,k_{z_{_{in}}}\,q_{z_{_{in}}} }{\left(\,k_{z_{_{in}}} +\,
q_{z_{_{in}}}\right)^{^{2}}}\,\frac{q_{z_{_{*}}}-k_{z_{_{*}}} }{q_{z_{_{*}}} +\,
k_{z_{_{*}}}}\,\,\exp\{\,i\,[\, q_{z_{_{*}}}\, a \sqrt{2} + (q_{z_{_{in}}} - k_{z_{_{in}}})\, (b-a)\,]\,\}\,\,.
\end{equation}
For $p$-polarized waves, we find
\begin{equation}
T_{_{\theta}}^{^{[p]}}(k_x,k_y)= \frac{4\,n^{\2}\,k_{z_{_{in}}}\,q_{z_{_{in}}} }{\left(\,n^{\2}k_{z_{_{in}}} +\,
q_{z_{_{in}}}\right)^{^{2}}}\,\frac{q_{z_{_{*}}}-n^{\2}k_{z_{_{*}}} }{q_{z_{_{*}}} +\,
n^{\2}k_{z_{_{*}}}}\,\,\exp\{\,i\,[\, q_{z_{_{*}}}\, a\sqrt{2} + (q_{z_{_{in}}} - k_{z_{_{in}}})\, (b-a)\,]\,\}\,\,.
\end{equation}
Due to the fact that the motion is on the $y$-$z$ plane, only second order $k_x$ contributions appear in the transmission coefficient, $T_{_{\theta}}^{^{[s,p]}}(k_x,k_y)$. Thus, without lost of generality, to calculate the complex amplitude $A_{_{out}}^{^{[s,p]}}(\boldsymbol{r},\theta)$  we can take the following approximation $T_{_{\theta}}^{^{[s,p]}}(k_x,k_y)
\approx T_{_{\theta}}^{^{[s,p]}}(0,k_y)$. Consequently,
\begin{eqnarray}
A_{_{out}}^{^{[s,p]}}(\boldsymbol{r},\theta) & \approx &   \frac{\mbox{w}_{\0}^{\2}}{4 \pi}\,\,
\int^{^{+\infty}}_{_{-\infty}}\hspace*{-.5cm}
\mbox{d}k_x\,\int^{^{+\infty}}_{_{-\infty}}\hspace*{-.5cm}
\mbox{d}k_y
\,\,T_{_{\theta}}^{^{[s,p]}}(0,k_y)\, \exp \left[-\frac{(k_{x}^{^2} +
k_{y}^{^2})\,\mbox{w}_{\0}^{\2}}{4}\,+\,i\,\varphi_{_{out}}(k_x,k_y;\boldsymbol{r},\theta)\,\right]
\nonumber \\
 & = &  \mathcal{G}\left[\,\frac{x}{\mbox{w}_{\0}}\,,\,\frac{\cos
(2\,\theta)\,y + \sin (2\,\theta)\,z}{z_{_{R}}}\,\right]\,\,  \mathcal{F}_{_{\theta}}^{^{[s,p]}}(y,z)\,\,,
\end{eqnarray}
where
\[ \mathcal{F}_{_{\theta}}^{^{[s,p]}}(y,z) =
\frac{\mbox{w}_{\0}}{2\, \sqrt{\pi}}\,\,\int^{^{+\infty}}_{_{-\infty}}\hspace*{-.5cm}
\mbox{d}k_y
\,\,T_{_{\theta}}^{^{[s,p]}}(0,k_y)\, \exp \left[-\frac{
k_{y}^{^2}\,\mbox{w}_{\0}^{\2}}{4}\,+\,i\,\varphi_{_{out}}(0,k_y;\boldsymbol{r},\theta)\,\right]\,\,.
\]
The detailed analysis of $\mathcal{F}_{_{\theta}}^{^{[s,p]}}(y,z)$ will be the subject matter of the next section.


\section*{\normalsize III. ASYMMETRIC GH EFFECT AND MULTIPLE PEAKS INTERFERENCE}
Let us consider the momentum distribution
\begin{equation}
g_{_{T}}^{^{[s,p]}}(k_x,k_y)= T_{_{\theta}}^{^{[s,p]}}(0,k_y)\,\, \exp \left[-\frac{(k_{x}^{^2} +
k_{y}^{^2})\,\mbox{w}_{\0}^{\2}}{4}\,\right]
\end{equation}
responsible for the shape of the transmitted beam. The contour plots of  $g_{_{T}}(k_x,k_y)^{^{[s,p]}}$  clearly show that for decreasing value of $k\mbox{w}_{\0}$, see Fig.\,2, and  for incidence angles approaching to the critical angle, see Fig.\,3,  the symmetry between $k_x$ and $k_y$ in the wave number distribution is broken. As anticipated in the Introduction, this {\em symmetry breaking} is responsible for the creation of a {\em dynamical} maximum. To examine in detail this phenomenon, let us first consider incidence at $\theta=\pi/4$,
\begin{equation}
\mathcal{F}_{_{\frac{\pi}{4}}}^{^{[s,p]}}(y,z)  =  \frac{\mbox{w}_{\0}}{2\, \sqrt{\pi}}\,\,
\int^{^{+\infty}}_{_{-\infty}}\hspace*{-.5cm}
\mbox{d}k_y
\,\,T_{_{\frac{\pi}{4}}}^{^{[s,p]}}(0,k_y)\, \exp \left[-\frac{
k_{y}^{^2}\,\mbox{w}_{\0}^{\2}}{4}\,-\,i\,
k_y\,y - \,i\,\frac{k_{y}^{^2}}{2\,k}\,z\,\right] \,\,.
\end{equation}
To estimate the maximum, we can apply the stationary phase method\cite{dingle},
\[
0=\left\{\frac{\partial}{\partial k_y} \left[\,\mbox{phase}\left[ T_{_{\frac{\pi}{4}}}^{^{[s,p]}}(0,k_y)  \right] - \, k_y\,y\,-\,\frac{k_{y}^{^2}}{2\,k}\,z\,\right]\right\}_{_{ky=\langle ky\rangle}}\,\,.
\]
Due to the fact that the phase of the transmission coefficient $T_{_{\theta}}^{^{[s,p]}}(0,k_y) $ is not dependent on the spatial coordinates, we can immediately find an analytical expression for the shift in  $y$ between  two maxima, i.e.
\begin{equation}
\Delta y = -\,\frac{\langle k_{y}\rangle}{k}\,\Delta z\,\,.
\end{equation}
For $\theta=\pi/4$, $n=\sqrt{2}$, and $k\mbox{w}_{\0}\geq 10$, the wave number distribution
is a symmetric distribution centered at $k_y=0$. Consequently, $\langle k_{y}\rangle=0$ and the maximum does {\em not} change its position. We thus recognize a {\em stationary} maximum. The numerical analysis confirms this theoretical prediction, see Fig.\,4.

Let us now consider incidence at critical angle,
\[ \sin\theta_c + \sqrt{n^{\2} -\sin^{\2}\theta_c} =\sqrt{2}\,\,.\]
For $n=\sqrt{2}$ the critical angle is $\theta_c=0$ and the transversal $y$-$z$ profile is determined by
 \begin{equation}
\mathcal{F}_{_{0}}^{^{[s,p]}}(y,z)  =  \frac{\mbox{w}_{\0}}{2\, \sqrt{\pi}}\,\,
\int^{^{+\infty}}_{_{-\infty}}\hspace*{-.5cm}
\mbox{d}k_y
\,\,T_{_{0}}^{^{[s,p]}}(0,k_y)\, \exp \left[-\frac{
k_{y}^{^2}\,\mbox{w}_{\0}^{\2}}{4}\,+\,i\,
k_y\,z -\,i\, \frac{k_{y}^{^2}}{2\,k}\,y\,\right]\,\,.
\end{equation}
In this case, the $z$-shift of the maximum in terms of the $y$ location of the detector is given by
\begin{equation}
\Delta z = \frac{\langle k_{y}\rangle}{k}\,\Delta y\,\,.
\end{equation}
For $k\mbox{w}_{\0}=10^{^3}$, the wave number distribution is not completely  symmetric in $k_y$ and this produce the first modifications  on the transmitted beam, see Fig.\,5.  Such a little modification is more evident for $p$-polarized waves. By decreasing the value of $k\mbox{w}_{\0}$ up to $10$, we lose the symmetry (see Fig.\,2) and  we clearly find a {\em dynamical} maximum. To estimate this dynamical shift, we observe that, as seen in section I,  for $k\mbox{w}_{\0}=10$ only positive values of $k_y$ contribute to the mean value, this implies
\begin{equation}
\langle k_y \rangle = \frac{2}{\mbox{w}_{\0}\sqrt{\pi}}
\end{equation}
 and, consequently, the shift in $z$ of the transmitted optical beam is given by
 \begin{equation}
 \label{law}
\Delta z =   \frac{2}{k\mbox{w}_{\0}\sqrt{\pi}} \,\Delta y\,\,.
\end{equation}
 The numerical analysis, shown in Fig.\,6, confirms such a prediction. In such a plot, it is also clear the {\em asymmetric} interference which appears in the presence of dynamical maxima.


\section*{\normalsize IV. CONCLUSIONS}

Optics sure represents a very stimulating field to reproduce quantum mechanical phenomena. For example, the well known Goos-H\"anchen shift\cite{GH} is the optical analogous of the {\em delay} time in non-relativistic quantum mechanics\cite{cohen,GH1}. These optical and quantum effects are due to the fact that evanescent waves  exist in the classical forbidden region.  This intriguing shift which is always matter of scientific investigation\cite{GH4,GH5,GH6,GH7} is, in general, a {\em stationary} shift. In this paper, we have analyzed in which situations this stationary shift becomes a {\em dynamical} shift.

Due to the fact that the dynamical shift is a direct consequence of the  breaking of symmetry in the wave number distribution, this new optical phenomenon can be also seen as an {\em asymmetric} GH effect. In our study,  we have seen that the more convenient circumstances to reproduce this {\em new} phenomenon are the choice of incidence at critical angles and of beam waists, $\mbox{w}_{\0}\approx 10/k \approx 1.6 \,\lambda$, of the order of the wavelength of the incoming gaussian beam.  This seems to be too restrictive for a possible experimental implementation of the theoretical analysis presented in this article. Nevertheless, this difficulty is very similar to the  difficulty found in detecting the standard Goos-H\"anchen sfhit which is of the order of the wavelength of the incoming beam. Consequently, it can be overcome with the same trick, i.e. {\em amplifying} the shift. For example, by preparing a dielectric structure which allows $2N+1$ internal reflections, we obtain for the transmission coefficient the following expressions
\begin{equation}
\left|\,T_{_\theta}^{^{[s,2N+1]}}(k_x,k_y)\,\right|= \frac{4\,k_{z_{_{in}}}\,q_{z_{_{in}}} }{\left(\,k_{z_{_{in}}} +\,
q_{z_{_{in}}}\right)^{^{2}}}\,\,\left|\, \frac{q_{z_{_{*}}}-k_{z_{_{*}}} }{q_{z_{_{*}}} +\,
k_{z_{_{*}}}}\,\right|^{^{2N+1}}
\end{equation}
and
\begin{equation}
\left|\,T_{_\theta}^{^{[p,2N+1]}}(k_x,k_y)\,\right|= \frac{4\,n^{\2}\,k_{z_{_{in}}}\,q_{z_{_{in}}} }{\left(\,n^{\2}k_{z_{_{in}}} +\,
q_{z_{_{in}}}\right)^{^{2}}}\,\,\left|\,\frac{q_{z_{_{*}}}-n^{\2}k_{z_{_{*}}} }{q_{z_{_{*}}} +\,n^{\2}k_{z_{_{*}}}}\,\right|^{^{2N+1}}\,\,.
\end{equation}
At critical angles, we have
\[ k_{z_{_{*}}}^{^{2}}>0 \,\,\,\,\,\mbox{for}\,\,\,k_y<0\,\,\,\,\,\,\,\,\,\,\,\,\mbox{and}\,\,\,\,\,\,\,\,\,\,\,\,
k_{z_{_{*}}}^{^{2}}<0 \,\,\,\,\,\mbox{for}\,\,\,k_y>0\,\,.
\]
Consequently,  by increasing the number of internal reflections, we can select the positive $k_y$ components in the transmitted wave number distribution for value of the beam waist, $\mbox{w}_{\0}$, greater than the wavelength, $\lambda$,  of the incoming laser beam.
The symmetry breaking in the wave number distribution, responsible for recovering the second order $k_y$ contribution to the phase which contributes to the maximum with the term $\langle k_y\rangle\,y/k$, can be thus optimized for experimental proposals by using the number of internal reflection $N$ and the ratio $\mbox{w}_{\0}/\lambda$.

 In a forthcoming paper, we shall analyze the asymmetric GH effect  for frustrated total internal reflection\cite{FTIR,FTIR2} and resonant photonic tunneling\cite{RPT}. Another interesting future investigation is represented by the possibility to include in our calculation the  focal shift\cite{GHc1}. This additional shift  represents a second order correction to the GH shift and consequently acts as a {\em delay} in the spreading of the outgoing optical beam.

\vspace{.8cm}

\noindent
\textbf{\footnotesize ACKNOWLEDGEMENTS}\\
 We gratefully thank  the Capes (M.\,P.\,A.), Fapesp (S.\,A.\,C.), and CNPq (S.\,D.\,L.)   for  the financial support and the referee for his useful suggestion on the new title and for drawing our attention to references on the GH shift and, in particular, on the interesting second order correction which leads to the focal shift\cite{GHc1}.



\newpage

\begin{figure}
\vspace*{-3cm} \hspace*{-2.7cm}
\includegraphics[width=19.5cm, height=29cm, angle=0]{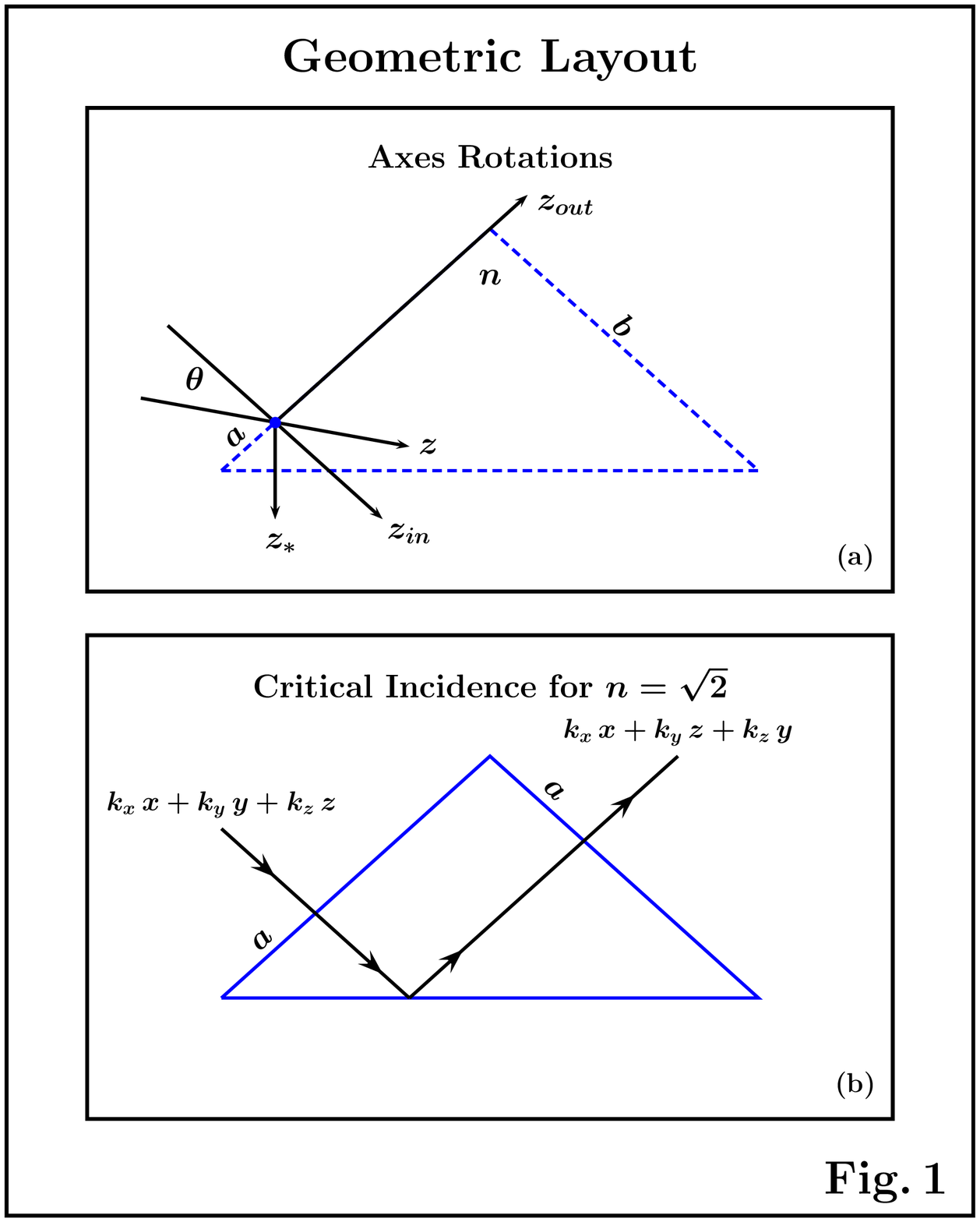}
\vspace*{-4.5cm}
 \caption{Geometric layout of the dielectric structure analyzed in this paper.}
\end{figure}

\newpage

\begin{figure}
\vspace*{-1.5cm} \hspace*{-2.3cm}
\includegraphics[width=19.5cm, height=27cm, angle=0]{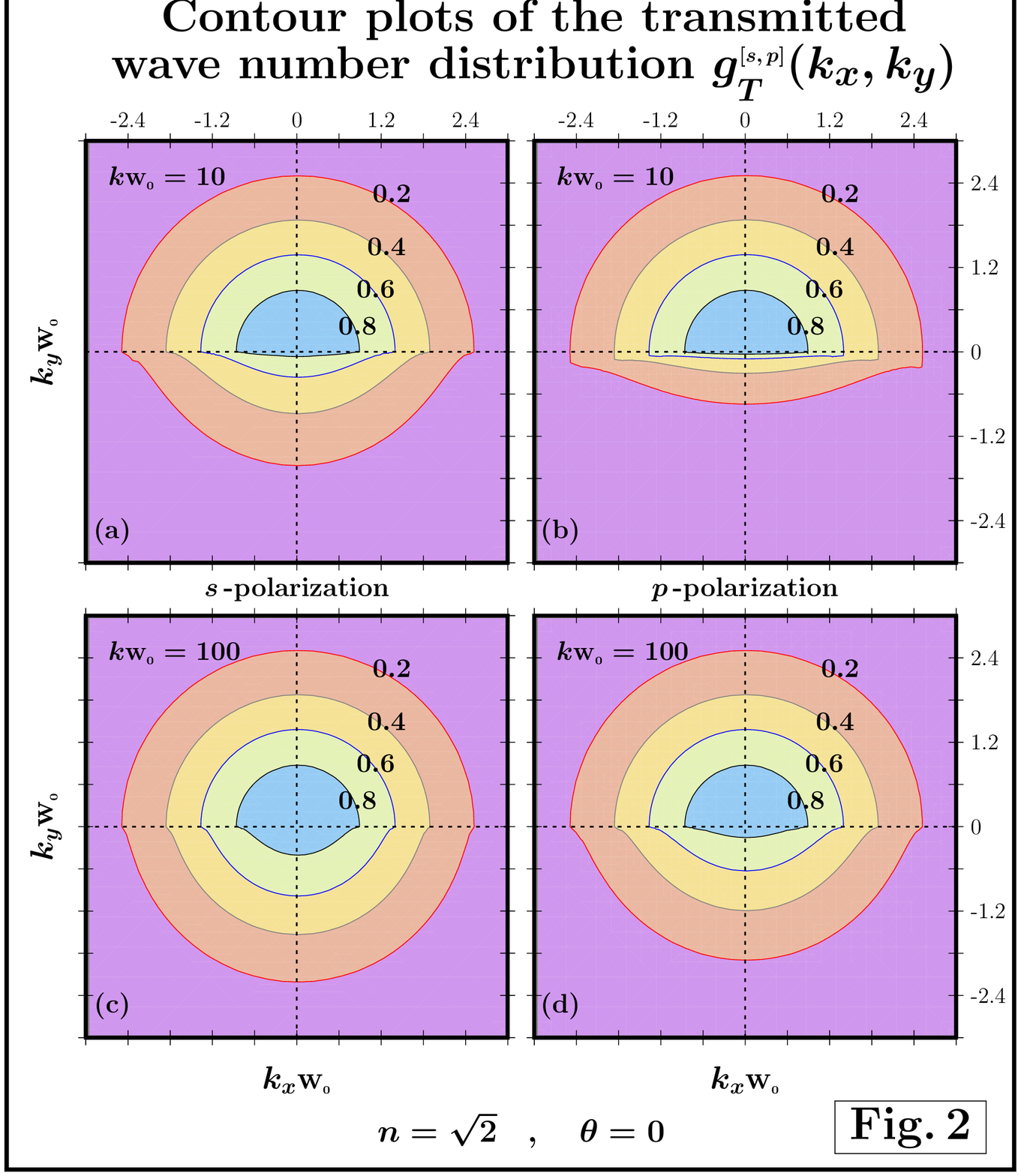}
\vspace*{-5.5cm}
 \caption{Contour plots of the transmitted wave number distribution, $g_{_{T}}(k_x,k_y)$, at critical angle for increasing values of $k\mbox{w}_{\0}$. The numerical data show that the symmetry, which is broken for $k\mbox{w}_{\0}=10$, is recovered by increasing the value of $k\mbox{w}_{\0}$ (total internal reflection).}
\end{figure}

\newpage

\begin{figure}
\vspace*{-1.5cm} \hspace*{-2.3cm}
\includegraphics[width=19.5cm, height=27cm, angle=0]{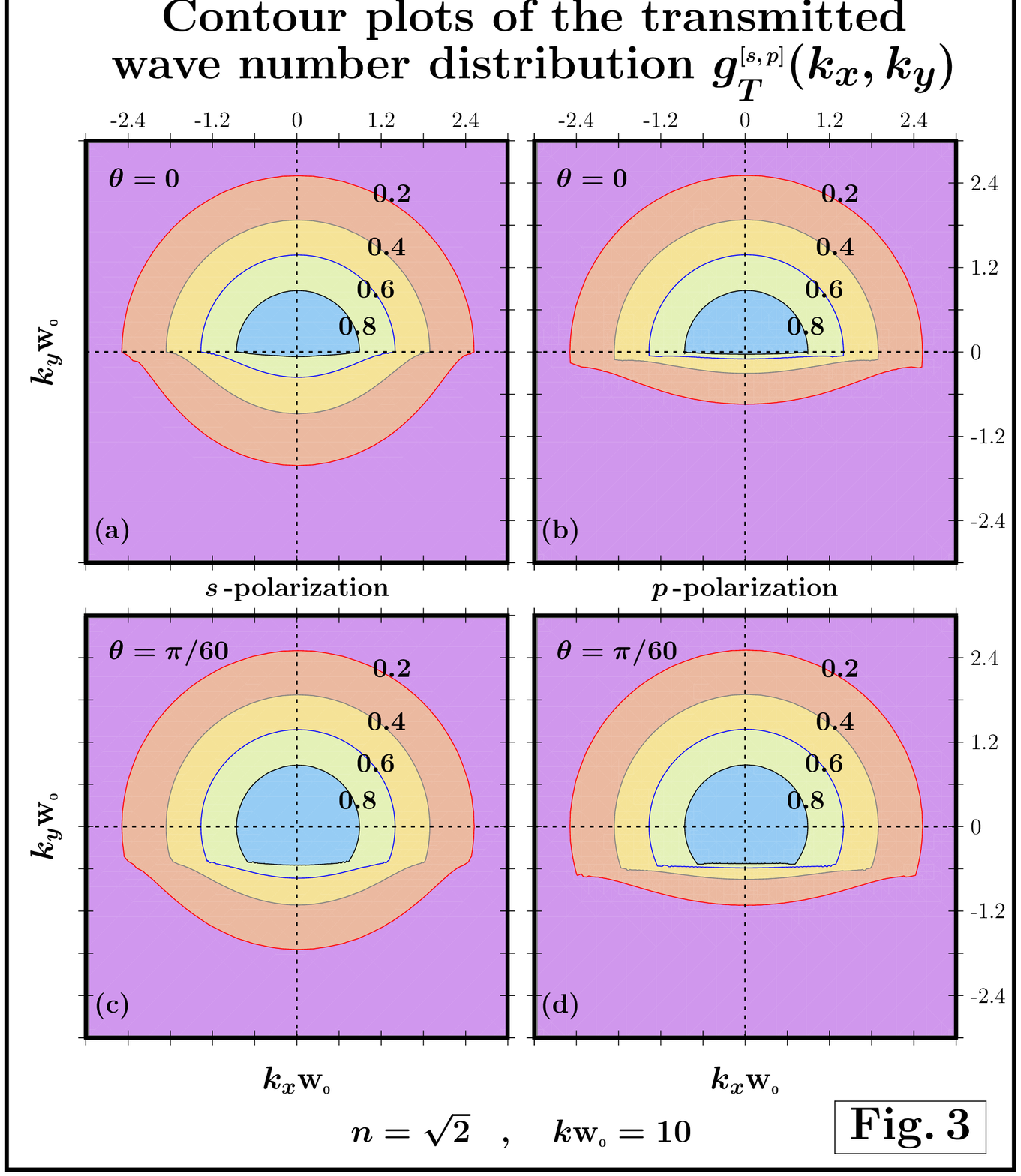}
\vspace*{-5.5cm}
 \caption{Contour plots of the transmitted wave number distribution, $g_{_{T}}(k_x,k_y)$, for $k\mbox{w}_{\0}=10$ and for increasing values of the incidence angle. The numerical data show that the symmetry, which is broken for $\theta=0$, is recovered by increasing the value of $\theta$ (total internal reflection). }
\end{figure}

\newpage

\begin{figure}
\vspace*{-1.5cm} \hspace*{-2.8cm}
\includegraphics[width=19.5cm, height=27cm, angle=0]{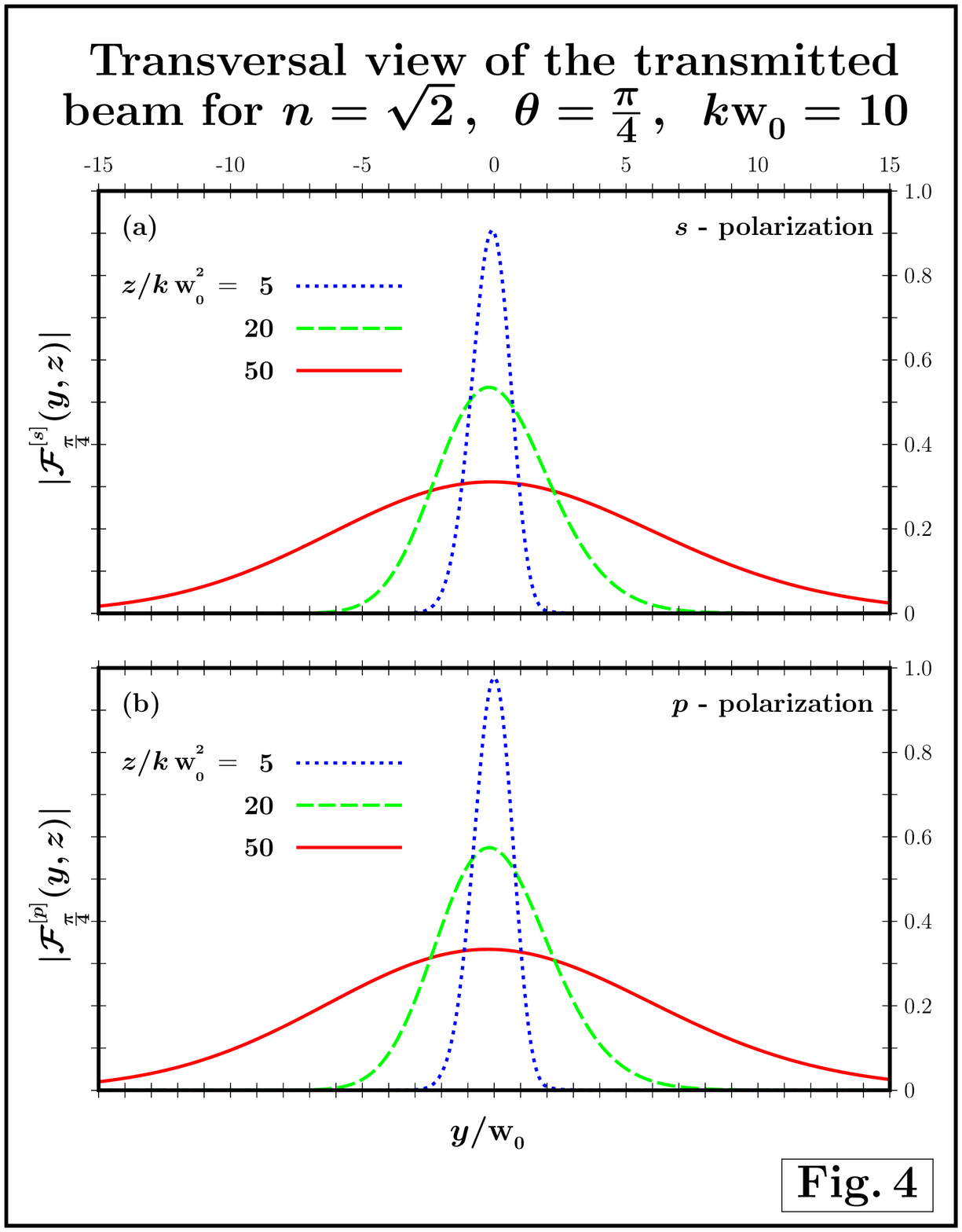}
\vspace*{-4.5cm}
 \caption{Transversal view of the transmitted beam for $s$ and $p$ polarized waves, $\mathcal{F}_{_{\frac{\pi}{4}}}^{^{[s,p]}}(y,z)$. Due to the symmetry of the transmitted wave number distribution, we find {\em stationary} maxima.}
\end{figure}

\newpage

\begin{figure}
\vspace*{-1.5cm} \hspace*{-2.8cm}
\includegraphics[width=19.5cm, height=27cm, angle=0]{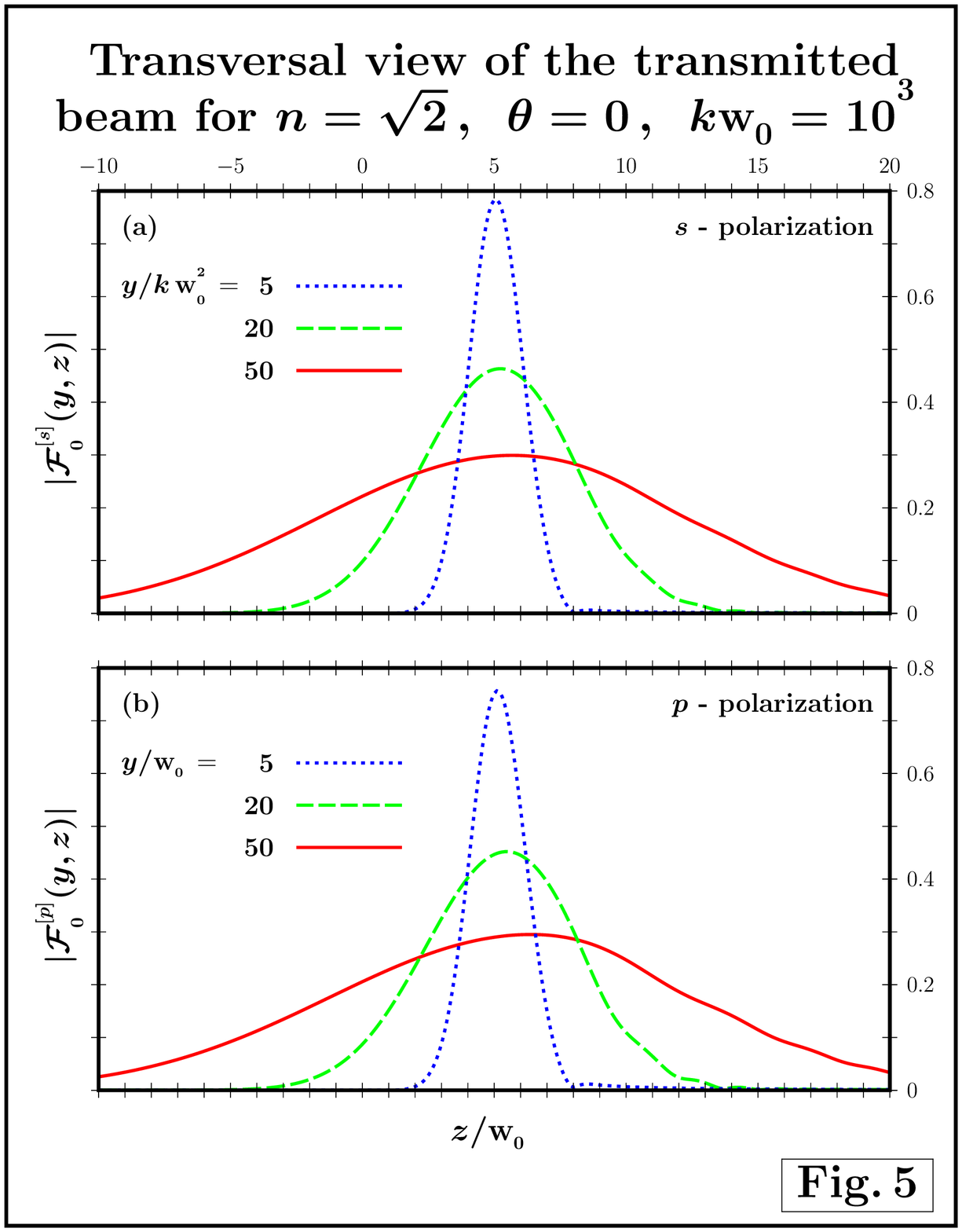}
\vspace*{-4.5cm}
 \caption{Transversal view of the transmitted beam for $s$ and $p$ polarized waves at critical angle, $\mathcal{F}_{_{0}}^{^{[s,p]}}(y,z)$, for $k\mbox{w}_{\0}=10^{^3}$. Due to the {\em partial} breaking of symmetry of the transmitted wave number distribution, we see the first modifications on the transmitted beam.}
\end{figure}

\newpage

\begin{figure}
\vspace*{-1.5cm} \hspace*{-2.8cm}
\includegraphics[width=19.5cm, height=27cm, angle=0]{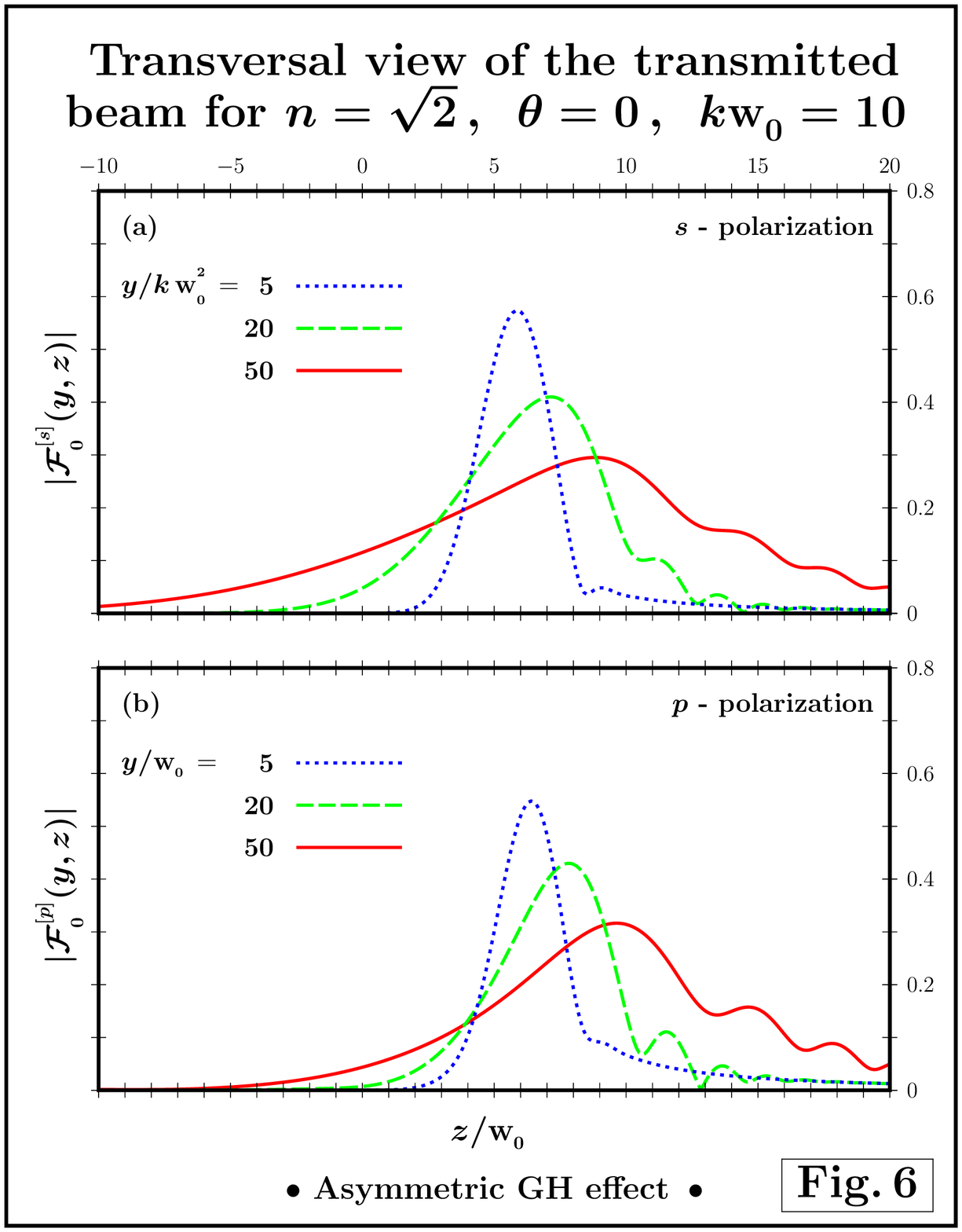}
\vspace*{-4.5cm}
 \caption{Transversal view of the transmitted beam for $s$ and $p$ polarized waves at critical angle, $\mathcal{F}_{_{0}}^{^{[s,p]}}(y,z)$, for $k\mbox{w}_{\0}=10$. Due to the {\em total} breaking of symmetry of the transmitted wave number distribution, the phenomena of dynamical shift and asymmetric interference clearly appear.}
\end{figure}

\end{document}